# PAMAM6 dendrimers and DNA: pH dependent "beads-on-a-string" behavior revealed by small angle X-ray scattering

Rolf Dootz,[a] Adriana Cristina Toma[b] and Thomas Pfohl*[a, b]

[a] MPI for Dynamics and Self-Organization, Bunsenstrasse 10, 37073 Göttingen, Germany. E-mail: rdootz@googlemail.com
[b] University Basel, Chemistry Department, Klingelbergstrasse 80, 4056, Basel, Switzerland. Fax: 0041612673855; Tel: 0041612673845; E-mail: adriana.toma@unibas.ch; thomas.pfohl@unibas.ch

**DNA interactions with polycations are not only important for our understanding of chromatin compaction but also for characterizing DNA-binding proteins involved in transcription, replication and repair. DNA is known to form several types of liquid-crystalline phases depending, among other factors, on polycation structure and charge density. Theoretical studies and simulations have predicted the wrapping of DNA around spherical positively charged polycations. As a potential mimic of the histone octamer or other DNA wrapping proteins, poly(amido amine) generation 6 (PAMAM6) dendrimers have been chosen for our study. The self-assembly of DNA induced by PAMAM6 has been investigated using small angle X-ray scattering (SAXS) in order to reveal the assemblies' structure dependence on the pH of the environment and on dendrimers concentration. We demonstrate that at pH 8.5 dense phases are formed and characterized by a 2D-columnar hexagonal lattice which is transformed into a 3D hexagonal lattice with increasing dendrimer concentration (charge ratio N/P). Moreover, a systematic analysis of the scattering data collected from diluted samples at pH 8.5 and 5.5 have led us to propose a wrapping scenario of DNA around PAMAM6 at pH 5.5.**

## 1 Introduction

The packing pathway of DNA inside the nucleus of a cell is still matter of great debate. It is currently accepted that the first level of organization of the genetic information is the nucleosome core particle.[1] A constant length of ~ 146 base pairs (bp) of DNA interact electrostatically and wrap 1.7 times around a complex of basic proteins called histones forming a pearl like necklace,[2] which through a yet unknown mechanism aggregate to form chromosomes. In order to understand DNA compaction *in vivo*, *in vitro* studies with spherical particles (e.g. dendrimers of different generations, nanoparticles)[3, 4] or linear polymers (e.g. poly-L lysine, poly-L arginine, polyethyleneimine, protamine)[5, 6] as model systems have already been explored giving good insights so far. As a continuation of these experiments, we decided to investigate how the size and charge distribution of particles determine the local structure of histone-like molecule-induced DNA self-assembly, employing polyamidoamine (PAMAM6) generation 6 dendrimers as a model system. PAMAM dendrimers are cationic polymers synthesised by a two-step iterative reaction that results in concentric generations of units around a central initiator core, having specifically controlled size, molecular mass and electrical charge.[7] Moreover, these highly monodisperse polymers, are often referred to as "artificial proteins" and are applied as such in cancer therapy.[8] Making an approximation in the size and shape of dendrimers, important proteins and bioassemblies (e.g. haemoglobin or nucleosome core particles) can be mimicked.[9, 10] Although, most dendrimers are globular, they can be used as well to mimic non-globular proteins.[11] Like proteins, dendrimer conformation adapts to the solvent, pH and ionic strength. Molecular dynamics simulations predicted that PAMAM dendrimers have an extended conformation at low pH resulting from repulsions between primary and tertiary positively charged amines. At neutral pH the size of the dendrimer decreases probably due to hydrogen bonding between uncharged tertiary amines and positively charged surface amines. At high pH, dendrimers contract since the global charge approaches neutrality.[11]

DNA interaction with dendrimers of different generations[12, 13] has attracted a wide range of communities either by desire of developing new non-viral vectors for gene therapy or wanting to decipher chromatin organization with the help of synthetic histone models. Under different experimental conditions (e.g. bulk, on surfaces or single molecule measurements) the DNA condensation process by dendrimers has been established to depend to a great extent on the dendrimers generation.[13-17] Dendrimers cooperative binding has been proven in terms of the coexistence between free and bound DNA molecules, dendrimers preferring to bind to already formed complexes rather than to DNA alone.[15] The morphology of aggregates is time dependent[18] and can be controlled by choosing the dendrimer generation: toroidal particles obtained with low generations (e.g. G1 and G2) and globular particles observed with PAMAM G6.[13] The mechanism of interaction is reversible and based on pure electrostatics, DNA structure and function not being persistently transformed,[19] and thus is of crucial importance for gene therapy applications. Recent simulations have shown that the dendrimer contracts during the interaction with DNA, the authors supposed this change came from a difference in the electrostatic environment of the polycation.[20] Modeling of these interactions helped in understanding that counterions and solvents contribute as well to the conformation of dendrimers.[21] In particular, DNA penetrates inside the dendrimers at neutral pH, while less penetration was reported at low pH, suggesting a better release from the complex when considering DNA delivery.[22] Simulating the interaction of dendrimers with several chain stiffness polyelectrolytes lead to the proposition of an optimum stiffness for the delivery and release from the dendriplex.[23] Since one of the most promising application of dendriplexes is gene transfer, Ainalem et. al. investigated the interaction between dendrimers/DNA aggregates and model biomembranes.[24] They concluded that aggregate morphology and charge play a role in the interaction with zwitterionic phospholipid bilayers. Moreover, their results suggest that lower generation dendrimer/DNA complexes have higher transfection potential than high generation ones.[24] Recently, PAMAM dendrimer/DNA complexes have been encapsulated in a polymer film degradable through surface erosion for gene delivery.[25] It is as well of high importance to understand DNA compaction by dendrimers in terms of gene expression dependence on the condensed state.[26]

Since dendrimers were proven to have good targeting efficiency, the structure of dendrimer induced DNA packing must be characterized and understood. From a theoretical point of view studies have been dedicated to the characterization of complexes formed between a positively charged sphere and a linear polyelectrolyte, mainly applicable to the 10 nm chromatin fiber. Netz and Joanny described the theoretical conditions for DNA wrapping and concluded that physiological salt concentrations are desirable.[27] Nguyen and Shklovskii published a theoretical phase diagram of the DNA self-assembly with positively charged spheres, and they have shown how the formation of these assemblies was dependent on charge ratio for several DNA concentrations.[28] Kunze and Netz treated a simple model case of DNA wrapped around histones, forming NCPs, showing how the wrapping interval is influenced by the ionic strength of the solution.[29] Beside the theoretical predictions of the structure of charged spheres with linear polyanions, X-ray diffractions analyses have been performed on systems containing DNA and proteins (NCPs) as well as on dendrimers and DNA. Phase diagrams have been built of NCPs aggregated by monovalent and multivalent (e.g. spermidine$^{3+}$, $Mg^{2+}$) cations.[30,31] X-ray diffraction structural data of these NCPs aggregates revealed a variety of liquid crystalline phases as a function of ionic strength and osmotic pressure (e.g. lamella-columnar phase, 3D orthoromibc crystals or 2D columnar hexagonal phase).[32] Fewer structural studies have been realized in order to give insight into the structure of dendrimer/DNA complexes, especially of those obtained with high generation dendrimers. Evans et. al. built the complete phase diagram of generation 4 dendrimers with DNA as a function of the ionic strength and charge ratio, revealing different liquid crystalline columnar phases.[33] Moreover, it was reported recently about the roll played by dendrimer genartion and ionic strenght on DNA containing aggregates morphology.[34]

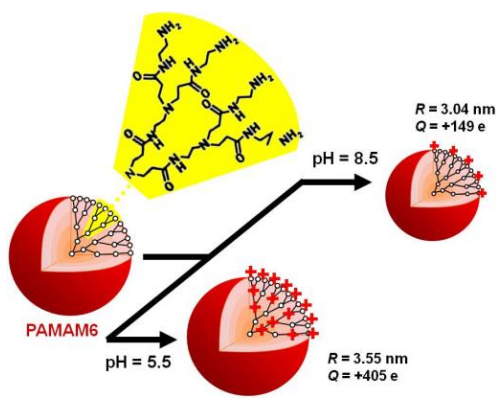

**Figure 1.** Schematic representation of PAMAM6 size and charge pH dependence.

Furthermore, the dynamics of self-assembly of DNA with G4 dendrimers has been monitored in microchannels through diffusive mixing, as a function of the charge ratio, and a transition from a columnar mesophase to a quadratic one has been reported.[35] More recently, Peng et. al. prepared and studied the structure of assemblies formed between DNA and PAMAM G4 with different degree of protonation. Increasing the degree of protonation the packing state of DNA was found to change from a squared lattice, further into a hexagonal lattice and probably organized as beads on a string at the highest degree of protonation.[36] For further reading on DNA self-assembly with different generations dendrimers and in different buffer conditions, we refer the reader to a recent review on this topic.[37]

DNA/dendrimer assemblies are much less studied and understood in comparison to the analogus DNA-histone complex.

We present in this article a detailed quantitative structural analysis of PAMAM6 cationic dendrimers induced nucleic-acid self-assembly - likely a crucial step in better understanding the first step of chromatin organization (the NCP) and how DNA interacts (wraps) with (around) spherical cations. In the present study, we used small angle X ray scattering (SAXS) to investigate how the structure of the formed assemblies is affected by several variable parameters (e.g. pH of the solution, charge ratio N/P). From our results we propose a wrapping model of DNA molecules around PAMAM6 dendrimers in a pH dependant manner. We compare our results at pH 5.5 to those of NCPs solutions and conclude that these environmental conditions are best fit for DNA wrapping. At high pH conditions we report precisely ordered 3D-crystalline structures with a remarkable degree of long-ranged ordering.

## 2    Materials and Methods

**Polymers and Sample preparation**

PAMAM6 and lyophilized highly-polymerized calf-thymus DNA were purchased from Sigma-Aldrich. pH-dependent PAMAM6 properties were determined in separate experiments.[38] The DNA concentration was keept constant at 5 mg/ml and the dendrimer concentration was varied. Both components were solubilised in 18.2 MΩcm water. The pH of all solutions was adjusted by adding $HCl_{aq}$ and $NaOH_{aq}$, respectively. After mixing, each sample was completely transferred to a quartz capillary (~1.0 mm diameter), and sealed.

**PAMAM6 denrimers and pH**

Figure 1 shows a schematic illustration of the pH-dependance of polyamidoamine dendrimers generation 6 (PAMAM6). When disolved in water, the primary amines of the outer PAMAM6 layers are partially protonated leading to a pH of 8.5.[39] At pH 5.5 – the pH of pure DNA solutions, all primary PAMAM6 amino-groups are protonated and tertiary amino-groups – situated in the dendrimer interior – are partially protonated. Thus, reducing solution-pH from 8.5 to 5.5 results in a 2.7-fold increase of dendrimer valence. As a consequence of the enhanced resulting intra-polymeric electrostatic repulsions, dendrimer branches are further extended leading to an increase in dendrimer radius of 16.7 %.[38]

**Small Angle X-ray Scattering (SAXS) and Data Treatment**

After mixing with DNA, PAMAM6/DNA complexes precipitate from solution. 2D small angle X-ray scattering (SAXS) is used to monitor directly the internal structure of PAMAM6/DNA assemblies at different compositions. Complex compositions are given in terms of the relative charge ratio N/P. N is the total number of amine groups of PAMAM6 and P is the total number of negatively charged DNA phosphate groups.[35] Due to their significantly higher electron density, the organisation of DNA molecules dominates the X-ray signal.

SAXS experiments were performed at $\lambda = 0.15498$ nm using a Bruker-AXS Nanostar with a virtually noise-free 2D-Hi-Star detector. All measurements were taken at ambient temperature with exposure times of 3600-7200 s. 2D-images were azimuthally averaged to produce 1D intensity profiles for scattering vectors $q \in [0.26$ nm$^{-1}$, $3.1$ nm$^{-1}]$ using the software Fit2D.

The scattering intensity $I(q)$ contains contributions from both the structure factor $S(q)$ and the form factor $F(q)$: $I(q) \sim S(q)F(q)$. $F(q)$ was determined from background-corrected SAXS measurements of sufficiently diluted solutions ($w/w \approx 0.5\%$). $S(q)$ was calculated by dividing the scattering intensity recorded from the precipitated phase by the corresponding $F(q)$ recorded in separate runs. Peak positions in $S(q)$ were determined by fitting with a Lorentzian shaped function.

The program GNOM was used to evaluate the pair distance distribution function $p(r)$ from $F(q)$.[40] Experimental data were corrected for instrumental broadening by numerical desmearing with the measured beam cross-section profile. To prevent any effects of inter-particle interactions at small $q$ values, for samples at pH 5.5 the calculation of the pair distance distribution function $p(r)$ and the determination of the radius of gyration $R_g$ were performed using the range $q > 0.3$ nm$^{-1}$.

The scattering intensity $I(q)$ of an ideal solution of NCPs was calculated from crystallographic coordinates published by Harp et al.[41] (PDB-file 1eqz.pdb) using CRYSOL.[42]

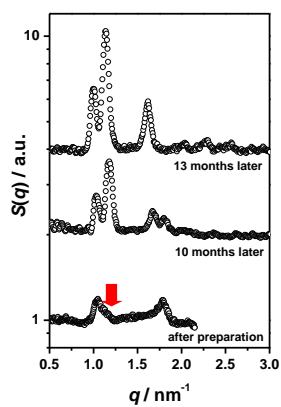

**Figure 2.** Evolution of PAMAM6/DNA assemblies structure with time. Structure factor represented at N/P=4.5 and pH 8.5 measured at several time intervals after preparation.

The program DAMMIN[43] was used to establish low-resolution bead-models of the 3D-shapes of PAMAM 6/DNA scattering entities in solution. Models are exploited *ab initio* from the experimental SAXS data without requiring any *a priori* structural information. Reconstructed shapes were obtained from averaging of at least six independent runs.

**Evolution of PAMAM6/DNA assemblies with time**

All X-ray patterns presented in Figure 3a-c were recorded approximately 13 months after sample preparation. The experimental conditions were designed to produce samples under equilibrium conditions in sealed capillaries. However, a significant evolution of complex structures is observable with equilibration time reflected in dramatic changes of the structure factor $S(q)$. Moreover, higher generation dendrimers were recently proposed to kinetically trap DNA into aggregates out of equilibrium.[14] Representative $S(q)$ obtained from PAMAM6/DNA complexes with a complex composition of $N/P = 4.5$ at pH 8.5 after different equilibration times are given in Figure 2. Immediately after preparation, $S(q)$ shows a single broad correlation peak at $q = 1.12$ nm$^{-1}$. When analysed with polarised optical microscopy, PAMAM6/DNA complexes reveal an optically (weak) birefringent pattern. This indicates that the DNA chains in the complex are orientationally ordered. It is hence reasonable to designate the observed mesomorphic structure as nematic liquid-crystalline. The packing lacks long-range positional order as the correlation peak is broad. From the ratio between the scattering vector and the full-width at half-maximum (FWHM) $q/\Delta q \approx 1.9$, it can be deduced that intermolecular positional ordering exists only between nearest neighbours. After several hours of equilibration time, $S(q)$ exhibits three sharp Bragg reflections at $q = 1.05$, 1.79, and 2.07 nm$^{-1}$. This indicates a 2D-columnar hexagonal organization with a lattice constant of $a_{hex,2D}= 6.9$ nm. The 2D-columnar hexagonal phase coexists with a nematic phase indicated by the broad peak at $q_{nem} = 1.51$ nm$^{-1}$.

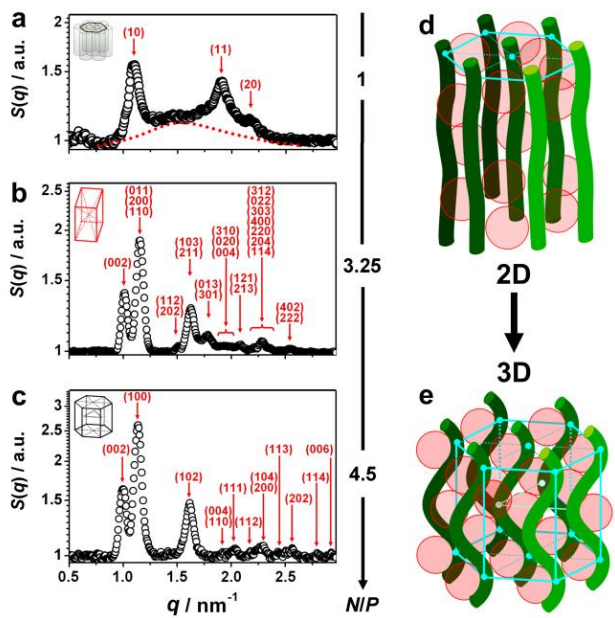

**Figure 3.** Structural organization of PAMAM6/DNA assemblies at pH 8.5. PAMAM6/DNA structure factors at a) N/P=1, 2D-columnar hexagonal phase; b) N/P=3.25, 3D-body-centered orthorhombic phase; c) N/P=4.5, 3D-hexagonal crystalline organization. A maximum amount of eleven sharp peaks are observed in the range of $q$ = 0.5-3.0nm$^{-1}$. All peaks are indexed according to the respective unit cell. In case of degenerated peaks (e.g. 100 and 010), only a single label is displayed for the sake of clarity. **d**, **e**, Schematic representations of the **d)** 2D to **e)** 3D structural transition of DNA (green) organisation induced by the optimum sized and charged PAMAM6 dendrimers (red).

However, there is an additional peak at $q$ = 1.13 nm$^{-1}$ with a significantly higher FWHM of $\Delta q$ = 0.21 nm$^{-1}$ (indicated by the red arrow in Figure 2). The peak at $q$ = 1.05 nm$^{-1}$ corresponds to the average distance between columns, whereas the peak at $q_c$ = 1.13 nm$^{-1}$ may correspond to the average spacing between PAMAM6/DNA scattering entities in a column. Accordingly, the reflection $q_c$ may be indicative of the onset of an ordered 3D organization of PAMAM6/DNA assemblies. After 10 months of equilibration time, the diffraction pattern is completely different exhibiting six Bragg reflections. The positions of these reflections are incompatible with both 2D- and 3D-hexagonal organisation. $S(q)$ rather corresponds to diffraction patterns obtained from a distorted 3D-body-centred orthorhombic organisation of PAMAM6/DNA scattering entities. The highly ordered 3D-hexagonal phase is fully established 13 months after sample preparation. From the data presented in Figure 2, it is not possible to be certain that the evolution of PAMAM6/DNA structure formation is complete after 13 months. It seems more likely that equilibrium values have not been reached yet. Despite possible changes of lattice constants, the highly ordered 3D-hexagonal lattice symmetry is expected to be preserved. This is consistent with structural organization time scales known from nucleosome core particle (NCP), where equilibration times of more than 17 months have been reported.[32]

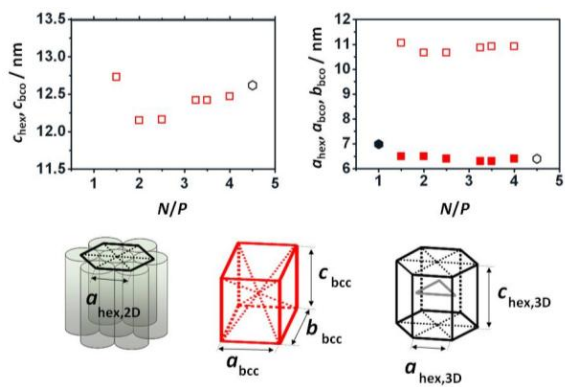

**Figure 4.** Assembly's composition dependence of lattice parameters indicate successive shifts in the molecular organization from 2D-columnar hexagonal (●) to 3D-hexagonal (○) via intermediate quasi-hexagonal (orthorhombic, □ and ■) states.

## 3 Results

An X-ray diffraction beam allowed us to measure the PAMAM6 induced DNA compaction dependence on the charge ratio for pH 8.5 and pH 5.5.

**Charge ratio influence on PAMAM6/DNA assemblies structure at pH 8.5**

The structure factor $S(q)$ of PAMAM6/DNA assemblies at $N/P = 1$ is presented in Figure 3a. $S(q)$ exhibits three small-angle Bragg reflections at $q = 1.103$, $1.907$, and $2.195$ nm$^{-1}$. These $q$-positions are in ratios close to $1:\sqrt{3}:\sqrt{4}$ suggesting a 2D-columnar hexagonal complex organization with a 2D-lattice constant of $a_{hex,2D} = 7.0$ nm. DNA molecules are suggested to be piled up regularly into columns that form a hexagonal lattice (Figure 3d). However, dendrimers are free to slide along columns exhibiting no intra-columnar correlations.

The presence of the broad reflection centred at $q_{nem} = 1.52$ nm$^{-1}$ can be interpreted in terms of a coexistence of the 2D-hexagonal phase with a nematic phase.[44]

The structural transition from a 2D-columnar hexagonal organisation at $N/P = 1$ to a 3D-hexagonal crystalline phase at $N/P = 4.5$ is taking place gradually via intermediate body-centred orthorhombic (bco) states. In Figure 3b, the structure factor recorded at $N/P = 3.25$ is shown. Observed peaks being best compatible to the space group $I2_12_12_1$ can be indexed within such a symmetry with unit-cell parameters $a_{bco} = 10.9$ nm, $b_{bco} = 6.3$ nm, and $c_{bco} = 12.4$ nm.[45] Figure 4 shows the $N/P$-dependence of lattice parameters. Continuous small variations indicate successive shifts in molecular organisation. The observed periodicity $c_{bco} = 12.7$ nm at $N/P = 1.5$ reflects the onset of intra-columnar correlations. With increasing $N/P$, the longitudinal order along the columns is further evolving and a 3D-quasi-hexagonal (orthorhombic) phase is fully established at $N/P = 2$. For all $N/P$, the ratio $a_{bco}/b_{bco} = 1.64$-$1.71$ is close to $\sqrt{3}$. This indicates that the 3D-hexagonal organization of PAMAM6/DNA entities is only slightly distorted. Independent of composition, particularly slow interaction kinetics are observed: establishing the highly ordered 3D-crystalline DNA organisation (for $N/P \geq 1.5$) takes place on remarkably long time scales of more than 12 months.

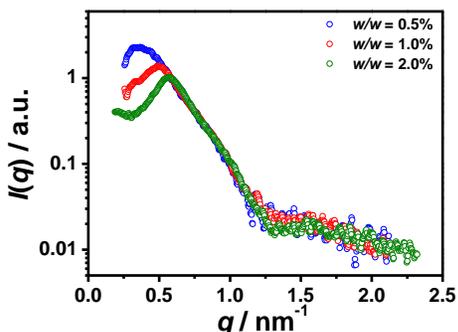

**Figure 5.** SAXS intensity profile from three aqueous solutions of PAMAM6/DNA of different mass fraction recorded at pH=5.5. All curves are scaled to the same intensity at the origin. A fixed complex composition of N/P = 1 is used for all three data sets.

In Figure 3c the structure factor of PAMAM6/DNA assemblies is represented at $N/P = 4.5$. The sufficient number of eleven Bragg reflections in the well-resolved X-ray diffraction pattern allows for an unambiguous identification of a 3D-hexagonal lattice (Figure 3e). Lattice constants are $a_{hex,3D} = 6.4$ nm and $c_{hex} = 12.6$ nm. The proposed lattice can account for a consistent indexation of all observed Bragg reflections. Moreover, no reflection expected from such a lattice is missing in the recorded diffraction pattern. The surprisingly large number of sharp Bragg reflections as well as the constant FWHM, $\Delta q$ of all diffraction peaks is indicative of crystalline-ordered domains with a high degree of positional order. The average size of a single-crystalline domain is estimated from the resolution-corrected FWHM of the first diffraction peak – $L_C = 2\pi/\Delta q_{(100)}$, $\Delta q_{(100)} \approx 0.01$ nm$^{-1}$ – to be ~630 nm (~$10^5$-$10^6$ unit-cells). In particular with regard to the enormous length and the polydispersity of the DNA, the strong 3D-long-range ordering is remarkable.

Structures of DNA condensates are strongly dependent on the correlation between compaction agent dimensions and effective charge. Accordingly, observed interactions of PAMAM6 and DNA are in strong contrast to those of DNA and smaller dendrimers. Compacted by smaller and less charged dendrimers, DNA molecules condense these along their chains forming liquid-crystalline mesophases with correlation lengths of 10-70 nm on time-scales of seconds.[44, 46] The 3D-hexagonal PAMAM6/DNA structure at pH 8.5 is rather corresponding to the structurally surprisingly similar 3D-orthorhombic organisation of NCPs in crystals and mesophases.[32, 47] In particular, the fact that both systems exhibit analogue equilibration times of several months is remarkable.[32] These similarities can be attributed to the fact that PAMAM6 and histone-cores possess comparable dimensions and charges. Due to the similar properties of the compaction agents, a (partial) wrapping of the DNA around PAMAM6 is expected. However, it is important to emphasize the fact that – contrary to NCPs, which consist of a well-defined amount of DNA wrapped onto histone octamers forming specific colloidal particles – PAMAM6/DNA complexes represent a real 3D-crystalline organisation of polydisperse DNA double-strands induced by optimum sized and charged PAMAM6 nanoparticles. Consistently, compared to NCP crystals, connecting (linker-)DNA strands are still present in the structure.

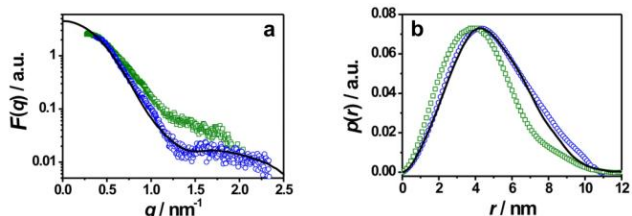

**Figure 6.** Determining shape and size of PAMAM6/DNA scattering entities. **a)** form factor $F(q)$ of PAMAM6/DNA scattering entities at pH 5.5(blue circles) and pH = 8.5 (green squares). A fixed assembly composition of $N/P = 1$ is used. Experimental data are contrasted to the scattering curve of an ideal NCP solution with no inter-particle interactions (black line) calculated from crystallographic coordinates published by Harp et al. (PDB-file 1eqz.pdb).[41] **b)** pair distance distribution function $p(r)$ of PAMAM6/DNA scattering entities at pH = 5.5 (blue circles) and pH = 8.5 (green squares) contrasted to $p(r)$ obtained from NCPs.

**Determining the form factor $F(q)$ of PAMAM6/DNA scattering entities.**

A reduction of pH to 5.5 is utilized to strongly increase the charge of DNA compaction agents. In order to elucidate the effects on the structure of PAMAM6/DNA scattering entities, PAMAM6/DNA form factors $F(q)$ are analysed. For sufficiently diluted samples, the scattering intensity is considered to be free from any detectable inter-particle effects.[48] SAXS measurements of such solutions may yield the form factor of the scattering entities. Figure 5 shows X-ray patterns of diluted aqueous PAMAM6/DNA solutions obtained at pH 5.5 and at fixed compostion of $N/P = 1$. For all measured mass fractions, $I(q)$ exhibits no Bragg reflections but a significant downturn for low-$q$ values. This reveals the existence of net repulsive interactions between PAMAM 6/DNA entities. With increasing PAMAM 6/DNA mass fraction, this downturn is more pronounced, directly reflecting the effect of stronger inter-particle interactions on the scattering intensity. Figure 5 shows that although inter-particle effects on $I(q)$ are still present at the lowest mass fraction of $w/w = 0.5$ %, they are already significantly reduced and shifted to lower $q$ values. Corresponding scattering data are used as an approximation of the (ideal) form factor. Consequently, to prevent any effects of the downturn at low $q$ values, the calculation of the pair distance distribution function $p(r)$ is performed using the range $q > 0.32$ nm$^{-1}$. To improve statistics in the high $q$ region, where inter-particle effects can reasonably be neglected even at higher sample concentrations, data obtained at $w/w = 0.5$ % are spliced with corresponding data of higher mass fraction ($w/w = 2.0$ %). The resulting form factor $F(q)$ is given in Figure 5a, and for comparison the corresponding form factor $F(q)$ recorded at pH 8.5, is also displayed. Furthermore, Figure 6b shows the corresponding calculated pair distance distribution functions $p(r)$. In particular, the pH-dependent slope of $F(q)$ at low $q$ values and the minimum around $q = 1.3$ nm$^{-1}$, which is observed only for the data recorded at pH 5.5, suggest a noticeable conformational modification of PAMAM6/DNA scattering entities with variation of pH-conditions. The increased maximum position of the bell-shaped $p(r)$ course and the significantly less pronounced tail at high-$r$ values indicate that scattering entities at pH 5.5 are larger and more compact. Moreover, in Figure 6a, PAMAM6/DNA data are contrasted to the scattering intensity of an ideal NCP solution with no inter-particle interactions, calculated from crystallographic coordinates published by Harp et al.[41] Most remarkably, PAMAM6/DNA at pH 5.5 and NCPs exhibit very similar form factors. Prominent common features are the superposition of $F(q)$ for $q < 0.8$ nm$^{-1}$ and the minimum at $q = 1.3$ nm$^{-1}$ and 1.4 nm$^{-1}$ for PAMAM6/DNA and NCPs, respectively. As $p(r)$ corresponds to the distribution of distances $r$ between any two scattering elements within a particle, it also offers an alternative way of calculating the radius of gyration $R_g$.[49] Obtained values for the radii of gyration are 3.3 nm and 3.8 nm for pH 8.5 and 5.5, respectively. In particular, the value of $R_g$ at pH 5.5 is very close to the value of 3.7 nm obtained from the crystallographic NCPs coordinates.[41] The overwhelming consistency of PAMAM6/DNA properties at pH 5.5 and NCP properties confirms that both entities possess similar overall shapes and sizes. Consistent to the NCP structure, a DNA wrapping scenario is expected for PAMAM6/DNA conjugates.

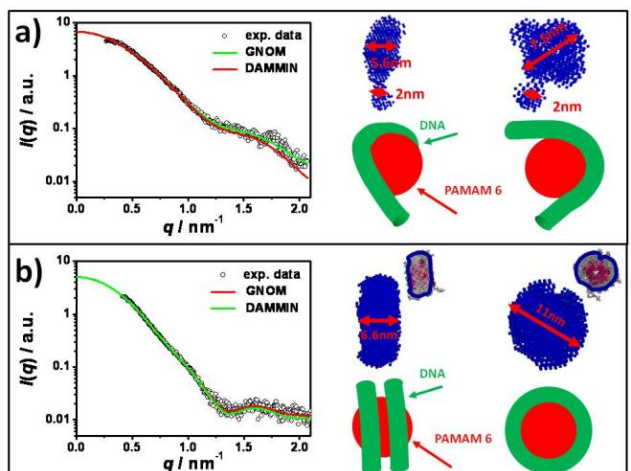

**Figure 7.** *Ab initio* model of PAMAM 6/DNA scattering entities at **a)** pH = 8.5 and **b)** pH 5.5. Below each model a schematic representation of the proposed structure is presented for PAMAM6/DNA scattering entities. At pH = 5.5, the structure of a NCP is given for comparison.

**Ab initio modelling.**

Small angle scattering data based modelling approaches allow for computation of low-resolution structural models of randomly oriented biological macromolecules in solution.[42] Based on $F(q)$ and $p(r)$, low-resolution structural models of the 3D-shapes of PAMAM6/DNA entities in solution are computed *ab initio*.[43] The starting point is a spherical search space of densely packed beads (dummy atoms), which completely encloses the particle. Simulated annealing is used as a global minimization algorithm that finally achieves a particle configuration matching the SAXS data. Annealing yields stable results for different, random starting points. The shape is recovered in an arbitrary orientation and handedness, the enantiomorph yielding the same scattering curve.[42] In order to further refine the solution an averaging of the results is used. This strategy is an established procedure to accumulate overall information and to achieve a sufficient degree of reliability because of the limited resolution of the SAXS experiments and the reliability of 3D-averaged information from specimens in solution.[44] Reconstructed shapes obtained from averaging of six independent runs are illustrated in Figure 7a. At both pH-conditions, obtained bead models match the recorded corresponding $I(q)$ data in the angular range $q < 1.4$ nm$^{-1}$. The program is not able to fit the high-$q$ portion of the experimental scattering pattern accurately (Figure 7a). However, it still provides a fair approximation of the overall appearance of the scattering entities.[33, 46] The most obvious structural feature of PAMAM6/DNA scattering entities at pH 8.5 is an ellipsoidal structure with a cross-section of approximately 7.5 nm and a height of about 5.6 nm (Figure 7a). Due to its dimensions and shape, the ellipsoidal structure is assigned to comprise the dendrimer ($2 \cdot R_{P6} \approx 5.6$ nm) and the tightly bound stretch of DNA ($d_{DNA} + 2 \cdot R_{P6} \approx 7.5$ nm). Due to their diameter of approximately 2 nm, the two smaller structures are assigned as unbound DNA stretches. The shape and location of the ellipsoidal structure comprising the dendrimer and the tightly bound stretches of DNA is well recovered in all runs. In Figure 7b, the bead model of PAMAM6/DNA entities at pH 5.5 shows a cylindrically shaped particle with almost circular cross-section. The particle diameter is 10.8 nm, whereas its height is 6.6 nm. Assigning again a $d_{DNA} = 2.0$ nm hard-core diameter to the DNA, the model suggests a PAMAM6 radius of 3.3 nm. Contrary to pH 8.5, the dendrimer seems to be completely encompassed by the DNA. Principle parameters derived from the bead model are listed in Table 1.

**Table 1.** Characteristic properties of PAMAM6 and PAMAM6/DNA derived from the bead model at pH 5.5 and 8.5

|  | pH = 5.5 | pH = 8.5 |
|---|---|---|
| **PAMAM 6** | | |
| radius in solution $R_{P6}$ | 3.55 nm | 3.04 nm |
| radius in complex $R_{P6C}$ | 3.3 nm | 2.8 nm |
| total charge $Q_{P6}$ | 405 e$^+$ | 160 e$^+$ |
| contribution of primary amino groups $Q_I$ | 256 e$^+$ | 160 e$^+$ |
| contribution of tertiary amino groups $Q_{III}$ | 149 e$^+$ | 0 e$^+$ |
| surface charge density $\Sigma_{P6}$ | 1.6-1.9 e$^+$/nm$^2$ | 1.46 e$^+$/nm$^2$ |
| **DNA** | | |
| absorbed amount of bp | 135 bp | 35 bp |
| absorbed length $L_a$ | 46 nm | 12 nm |
| diameter $d_{DNA}$ | 2 nm | |
| line charge density $\tau_{DNA}$ | 6 e$^-$/nm | |
| persistence length $L_p$[249] | 50 nm | |
| covered dendrimer surface area $A_{P6C}$ | 92 nm$^2$ | 24 nm$^2$ |
| absorbed charge $Q_{DNA}$ | 270 e$^-$ | 70 e$^-$ |
| local overcharging ($Q_{DNA}/A_{P6C}$-$\Sigma_{P6}$) | 1.0-1.3 e$^-$/nm$^2$ | 1.46 e$^-$/nm$^2$ |
| bending energy $E_b$ | 62 $kT$ | 22 $kT$ |

Schematic representations of the proposed structure of the PAMAM6/DNA scattering entities are given in Figure 7. Bead models yield positive proof of a DNA wrapping scenario. Although a multitude of slightly different conformations of single scattering entities may contribute to the form factor, the *ab initio* models indicate rather specific conformations. One may speculate that there are distinct energetic constrains that favour such a rather specific wrapping path of the DNA around PAMAM6. This is compatible with the observation of the highly ordered, 3D-superstructure of PAMAM6/DNA complexes at high concentrations.

The principle organization of the DNA chain on the PAMAM6 surface is primarily determined by the pH-dependent dendrimer valency. At pH 8.5, attractive electrostatic interactions of both components are not strong enough to completely overcome elastic and repulsive counterbalances and to induce a complete DNA wrapping. As a consequence, the PAMAM6 charge density of +1.46 e/nm$^2$ results in an adsorption of only 12 nm (~35 bp) of DNA in approximately half a turn. A charge density of about +1.6-1.9 e/nm$^2$ at pH 5.5 is needed to enable a complete wrapping of the DNA in approximately 1.7 turns, corresponding to 46 nm (~135 bp) of absorbed DNA. The result is a cylindrically shaped conformation of PAMAM6/DNA entities, which reproduces NCPs in structure and dimensions. In this sense, PAMAM6/DNA entities are excellent biomimetics of NCPs. Consequently, one may conclude that, at least for a simple replication of the beads-on-a-string structure without paying attention to DNA functionality, the specific arrangement of charged patches on the histone-octamer is not required.

## 4      Discussion

**Characteristic properties of PAMAM6 and PAMAM6/DNA derived from the model.** For a semi-flexible polyelectrolyte with a certain bending rigidity and therefore a finite persistence length $L_P$, the chain bending upon adsorption onto a sphere surface disfavours wrapping. Two effects are included in $L_P$, namely the mechanical bending rigidity of the uncharged chain and electrostatic repulsions of like-charged chain segments.[50, 51] With the parameters in Table 1 at hand, it is possible to estimate the bending penalty, which has to be overcome by adsorption energies in order to enable a wrapping of the determined stretches of DNA around the sphere. Obtained results show that an energy gain upon adsorption of approximately 62 $kT$ is needed in order to achieve a complete DNA wrapping in about two turns around the dendrimer. However, obtained values represent a lower estimate, since other contributions such as twist and torsion of the polyelectrolyte chain are neglected. At pH 5.5, the large amount of DNA adsorbed on the dendrimers leads to a strong, overall overcharging of the spherical cation. The overall charge of the DNA adsorbed on the PAMAM6 surface exceeds the initial charge of the dendrimer, so that the net charge of the entity changes sign. The phenomenon of overcharging of spherical macroions complexed with oppositely charged, linear polyelectrolytes is well known in literature.[52, 53] Consequently, these inter-particle repulsions prevent the organization of long-range ordered complex structures. It is important to point out, that at pH 8.5 overcharging effects are also present. However, they are only locally encountered, but are more pronounced (Table 1). Assigning a $d_{DNA}$ = 2.0 nm hard-core diameter to the DNA,[33] results presented in Figure 7 suggest a PAMAM6 radius of 2.8 nm and 3.3 nm at pH 8.5 and 5.5, respectively. However at corresponding solution pH, PAMAM6 dendrimers in solution possess a larger radius of 3.04 nm and 3.55 nm, respectively. This indicates that PAMAM6 shrink upon the interaction with oppositely charged DNA, in aggrement with recent theoretical predictions.[54]

The experimentally observed transition from an adsorption of a short stretch of DNA to a full wrapping with increasing the valence of the almost spherical macroions is in good agreement with theoretical studies on the interaction of a spherical macroion and a linear, oppositely charged polyelectrolyte in dilute solutions.[29] Contrary to all theoretical studies, which focus on the interplay of only very few components, the experimentally analysed PAMAM6/DNA system comprises complex interactions of a multitude of nanoparticles and DNA strands.

Structural studies on dendrimer induced DNA self-assembly reported predominantly on hexagonal or square lattice dense phases depending on dendrimer size and solution conditions. Lower generation dendrimers G2 and G3 form hexagonal and square lattices with DNA, respectively, mantaining a B-form of nucleic acids upon interaction.[55] The largest amount of scattering data can be found for assemblies formed with G4 dendrimers. For example, Evans et al. reported on a square to hexagonal lattice transition when increasing the charge ratio.[33] Furthermore, dynamic interaction experiments in microchannels (where kinetically trapped states are very much avoided) underscored again the presence of a quadratic lattice.[35, 46] Moreover, the recent experiments of Chen et al.[56] have investigated the effect on assemblies with G4 of the degree of dendrimer protonation at a fixed charge ratio. It was described that increasing the degree of protonation (decreasing the pH of the solution), assembly structure evolves from square and/or hexagonal lattices packing for low protonation (high pH) conditions; going through predominantly hexagonal ones for intermediate protonation degrees (neutral pH) and even proposed a beads-on-a-string like DNA wrapping (1.4 turns) around G4 for highly protonated samples (low pH). The DNA wrapping model suggested for the low pH range is in agreement with the results presented in this work, reinforcing the hypothesis of an acidic pH environment favoring the wrapping of a linear polyanion around a positively charged sphere.

Model systems of natural DNA wrapping not only shed light on some aspects of chromatin folding but also on other proteins or bio-assemblies around which DNA was found to wrap around.[57] DNA biological functions (e.g. replication, transcription, recombination and repair) strongly depend on the surface interactions between a variety of nucleic acid binding proteins which regulate DNA conformation. Models of DNA wrapping have already been proposed for *lac* represor, based on its structure.[58] Furthermore, a non-specific salt dependent DNA wrapping driven by coulombic interacions has been revealed on the surface of the *lac* repressor tetramer. A negative supercoil is induced into DNA upon its wrapping around a specific domain of DNA gyrase protein.[59] A large number of cationic side chains can be found at the wrapping interface, being a generality of DNA wrapping proteins with highly diverse functions (involved in regulation, topology and repair).[60] From transmision electron microscopy studies it was inferred that DNA wraps around cationic nanospheres in order to minimize the polyanion's bending rigidity.[4]

Our structural study may thus have important implications for other proteins (particles) around which DNA wraps, for example the ones envolved in DNA damage recognition.[61]

Since analyzing the complex interplay involved in the organization of a beads-on-a-string structure is far beyond the possibilities of simulations, experimental results obtained from the PAMAM6/DNA model-systems are expected to be an important contribution to the field.

The results presented here yield information about the formation of PAMAM6-assisted multidimensional structures of long-chain DNA double-strands. Controlling PAMAM6 valency and content allows for a defined electrostatic tuning of DNA structure dimensionality from 1D to 3D. Observed DNA structures are artificial equivalents to *in vivo* DNA organisation. Owing to the fact that connecting (linker-)DNA strands are still present in PAMAM6/DNA conjugates, we expect the PAMAM6/DNA system to represent an ideal starting point towards bio-engineering of an artificial 30 nm-fibre.

## 5 Conclusions

We have investigated the structure of dense phases resulting from the self-assembly of DNA with PAMAM6 dendrimers in pure water at variable pH values and N/P charge ratios. 2D-columnar hexagonal phases have been observed at pH 8.5 and low charge ratios. Increasing the dendrimer concentraion we first revealed the presence of body-centered orthorombic (bco) phases which evolved further on with the dendrimer concentration to 3D-hexagonal lattices. Analysing dilute solutions of PAMAM6/DNA at pH 8.5 and 5.5 and using standard simulations programs we propose two binding scenarios. From a bending energy cost point of view, the DNA chain has been found to wrap around PAMAM6 at pH 5.5 favorably, and a similarity with the NCPs or other DNA-wrapping proteins has been further proposed.

**Acknowledgements**

We gratefully acknowledge fruitful discussions with Heather Evans, Stephan Herminghaus and Sravanti Uppaluri. This work was supported by the Deutsche Forschungsgemeinschaft (PF 375/2); and Sonderforschungsbereich [755 "Nanoscale photonic imaging"].